\documentclass[twoside,leqno,twocolumn]{article}
\usepackage{ltexpprt}

\usepackage{amsmath}
\usepackage{amssymb}
\newcommand{\mapping}[2]{\langle #1 : #2 \rangle}

\usepackage{empheq}
\usepackage{times}

\usepackage{pifont}
\usepackage{epsfig}
\usepackage{tabularx}
\usepackage{amssymb}
\usepackage{latexsym}
\usepackage{amsmath}
\usepackage{delarray}
\usepackage{moreverb}
\usepackage{xspace}

\usepackage{lscape}
\usepackage{changebar}
\usepackage{graphicx}
\usepackage{graphics}
\usepackage{mflogo}
\usepackage{xspace}
\usepackage{texnames}
\usepackage{rotating}
\usepackage{alltt}

\usepackage{amsmath}
\usepackage{epsfig}
\usepackage{booktabs,paralist}

\newcommand{\N}{\mathbb{N}}

\newcommand{\singlefigure}[4]{{\begin{figure}[htp] %
\centering %
\includegraphics[width=#1\linewidth]{#2}\\
\caption{#3}%
\label{#4}%
\end{figure}}}

\begin{document}

\title{Mapping and Matching Algorithms: Data Mining by Adaptive Graphs}

\author{Paolo D'Alberto \thanks{pdalberto@ninthdecimal.com} 
  \and Veronica Milenkly \thanks{vmilenkiy@ninthdecimal.com}}

\maketitle

\begin{abstract}
\small\baselineskip=9pt Assume we have two bijective functions $U(x)$
and $M(x)$ with $M(x)\neq U(x)$ for all $x$ and $M,N: \N \rightarrow
\N$ . Every day and in different locations, we see the different
results of $U$ and $M$ without seeing $x$. We are not assured about
the time stamp nor the order within the day but at least the location
is fully defined. We want to find the matching between $U(x)$ and
$M(x)$ (i.e., we will not know $x$). We formulate this
problem as an adaptive graph mining: we develop the theory, the
solution, and the implementation. This work stems from a practical
problem thus our definitions. The solution is simple, clear, and the
implementation parallel and efficient. In our experience, the problem
and the solution are novel and we want to share our finding.
\end{abstract}

\section{Introduction}
\label{sec:introduction}

Let start by introducing the problem by its practical case. We are
traveling with our smart phone. We take a taxi and go to the
airport. We surf using our data plan. We arrive at the airport and we
connected to the local WiFi, we surf. Before boarding we turn off our
phone. We land and the previous process restarts. During our surfing,
our phone will be identified by a unique number as a function of the
device and application (i.e., UUID). While we are using the WiFi, our
device will have also a MAC address and IP. If we have the distinct
set of MACs and UUIDs, can we find the match: what UUID is associated
with the MAC?

If we identify our phone as $x$, we have two deterministic functions:
function $U(x,t,\ell)$ with location $\ell$ and time $t$ that
identifies our unique device, and function $M(x,t,\ell)$ with location
and time that identifies the MAC. We have only a sample in time of
$U(x,t,\ell)$ and a sample by location of $M$.  In practice, We may
not gain $U(x)$ and $M(x)$ at the same time but in a reasonable
interval of time, say one day: for example, at a specific airport and
date (day) we may have either one but not both with no specific time
information beside the day.  Also, given $x$ we may have $S =
U(x,t_i,\ell_j)$, that is $U(x)$ is not unique and it may the
composition of a set of exclusive functions $U_i(x)$ but when possible
we enforce a deterministic and unique result.

The problem boils down as to answer the following question: If we are
observing the output of $U$ and $M$, can we guess $x$, which is
associated to $U(x)$ and $M(x)$?

We define an airport as $L_i$ with $i\in [0,N-1]$. There are $N$
airports and we enumerate them. We describe the first day we observe
events simply as $t_0$. Thus $t_1$ is the second day: this will imply
that day $t_i$ precedes $t_{i+1}$.

Let us start considering the first day $t_0$.  For every $L_i$ there is a
set of associated MAC address, we identify this set as
$M_{t_0}^{L_i}$.  Also, we determine the users in one mile radius from
$L_i$: We identify this set as $S_{t_0}^{L_i}$.
\begin{equation}
  S_{t_0}^{L_i} = \{u : dist(u,L_i)<1\mbox{ at time }t_0\}
\end{equation}

The user set $S_{t_j}^{L_i}$ is not complete because we have only a
sample of the available impressions: we sample in time the values of
$U(x,t,\ell)$, we cannot keep an ordered time sequence beside a day
granularity, and by construction we may cover only a small area of the
airport $L_i$.

In practice, we associate $\mapping{S_{t_0}^{L_i}}{M_{t_0}^{L_i}}$,
the departing addresses to the departing users. This is the mapping we
would like to refine as much as possible, until we can have a
one-to-one matching. That is, we can infer the hidden $x$ that
determines the unique mapping between $U(x)$ and $M(x)$. These same
users are landing to different $L_j$ with $i\neq j$ and thus different
addresses $M_{t_0}^{L_j}$ and $M_{t_1}^{L_j}$ may be given.

Every mapping $\mapping{S_{t_j}^{L_i}}{M_{t_j}^{L_i}}$ describes a
graph, a fully connected bipartite graph. We need to combine all
mappings as above in order to achieve our goal. This is an adaptive
graph algorithm: we build a graph step by step, day by day. We check
whether mappings in different graphs have intersection and we can
split the graph by cutting edges.

The final goal is to grind these mappings into matches where one user
is associate to one MAC or at least to the finest refinement possible.
In the following, we formulate our solution using the same notations,
we present our algorithm, a few simplifications, and our results.

\section{The algorithm}
\label{sec:algorithm}
Consider the mappings for the first day:
$\mapping{S_{t_0}^{L_i}}{M_{t_0}^{L_i}}$. These can be considered as
the departing addresses for the departing users. The users departing
from $L_i$ can be the user landing at $L_j$ with $j\neq i$.  If there
is intersection between addresses we could refine the mapping:

{\tiny 
  \begin{equation}
    \begin{split}
      {\mapping{S_{t_0}^{L_i}}{M_{t_0}^{L_i}}} =
      & \mapping
      {S_{t_0}^{L_i}\smallsetminus \Big(\bigcup_{n=0,1}^{i \neq j} S_{t_n}^{L_j}\Big)}
      {M_{t_0}^{L_i} \smallsetminus \Big(\bigcup_{n=0,1}^{i\neq j}M_{t_n}^{L_j}\Big)} 
      + \\
      & \sum_{j\neq i} 
      \mapping
          {S_{t_0}^{L_i}\cap \big(\bigcup_{n=0,1}S_{t_n}^{L_j}\big)}
          {M_{t_0}^{L_i}\cap \big(\bigcup_{n=0,1}M_{t_n}^{L_j}\big)} \\
    \end{split}
    \label{eq:partition1}
  \end{equation}
} Here the $+$ operation is the disjoint concatenation of mappings
binding fewer elements and refining them towards matches. Our
interpretation of Equation \ref{eq:partition1} follows: if there is
any landing information the mapping between departing users and
departing address, then we can refined the mapping into two major
components.  {\tiny
  \begin{equation} 
    \mapping{\dot{S}_{t_0}^{L_i}}{\dot{M}^{L_i}_{t_0}} \equiv 
    \mapping {S_{t_0}^{L_i}\smallsetminus \Big(\bigcup_{n=0,1}^{i \neq j}
      S_{t_n}^{L_j}\Big)} {M_{t_0}^{L_i} \smallsetminus
      \Big(\bigcup_{n=0,1}^{i\neq j}M_{t_n}^{L_j}\Big)} 
    \label{eq:p1refined}
  \end{equation}
} Equation \ref{eq:p1refined} and \ref{eq:partition1}.(part one) refer
to the departing users without landing information.  { \tiny
  \begin{equation}
    \sum_j\mapping{S^{L_i\rightarrow L_j}_{t_0}}{M^{L_i\rightarrow
        L_j}_{t_0}}
    \equiv
    \sum_{j\neq i} \mapping{S_{t_0}^{L_i}\cap
      \Big(\bigcup_{n=0}^1 S_{t_n}^{L_j}\Big)} {M_{t_0}^{L_i}\cap
      \Big(\bigcup_{n=0}^1 M_{t_n}^{L_j}\Big)} 
    \label{eq:p2refined}
  \end{equation} 
}
If there is an intersection between departing and landing users, we
refine the mapping with the intersection of the departing and landing
addresses. The locations are disjoint, very likely a user will be only
at two locations in two days and thus $S_{t_0}^{L_i}\cap
\big(\cup_{n=0,1}S_{t_n}^{L_j}\big)$ will be true only for one $j$,
then the mappings are disjoint. Because we are considering two
consecutive days we may have to combine two or more mappings one step
further. Let us introduce the product of mappings.

By definition, if we have a $\mapping{A}{M}$ any user in $A$ can be
mapped to any address in $M$. As a graph, this represents a bipartite
fully-connected graph. If we have another mapping $\mapping{B}{N}$ and
there is an intersection between addresses, then we know that the same
users should be in both $A$ and $B$. After all the address is unique
to the device.  It makes sense to take the intersection of users as
well. In practice, we assume users will have likely or consistently
the same user identification number. This will refine the mapping
reducing the size of the three resulting mappings: \footnote{This
  definition does not fully represent reality. For example, a
  $M(x_0)=m_0$ is unique and $U(x_0)$ is not unique say $u_0$ in
  $\mapping{A}{M}$ and $u_1$ in $\mapping{B}{N}$, then in Equation
  \ref{eq:product} we have $\mapping{u_0}{\varnothing} +
  \mapping{\varnothing}{m_0} + \mapping{u_1}{\varnothing}$ =
  $\varnothing$ instead of $\mapping{u_0,u_1}{m_0}$. The definition of
  * operation is seeking for a deterministic and unique in time match.}
\begin{equation}
  \begin{split}
    \mapping{A}{M}*\mapping{B}{N} = & \mapping{A\smallsetminus
      B}{M\smallsetminus N} + \\ & \mapping{A\cap B}{M\cap N} + \\ &
    \mapping{B\smallsetminus A}{N\smallsetminus M} \\
  \end{split}
  \label{eq:product}
\end{equation}

We use the $*$ operator to represent this operation. In combination
with $+$ operator, our algorithm will be based an algebra. If there is
no intersection, there is no refinement and
$\mapping{A}{M}*\mapping{B}{N} = \mapping{A}{M} + \mapping{B}{N}$.

If there are only two mappings the product is intuitive and the final
result is a disjoint mapping. Let us consider two mappings composed by
disjoint simpler mappings and their products
\begin{equation}
{\cal P}= \Big(\sum_{j=0}^{K-1}w_j\Big)*\Big(\sum_{i=0}^{L-1}v_i\Big)
\end{equation}
where $w_j= \mapping{A_j}{M_j}$ and $v_i=\mapping{B_i}{N_i}$. We will
abuse the set notation a little here: 
\begin{equation}
  \begin{split}
    {\cal P}= &\sum_{j=0}^{K-1}\Big( \sum_{i=0}^{L-1} w_j\smallsetminus v_i\Big) + \\
    & \sum_{j=0}^{K-1}\sum_{i=0}^{L-1}w_j\cap v_i +\\
    & \sum_{i=0}^{L-1}\Big( \sum_{j=0}^{K-1} v_i\smallsetminus w_j\Big)\\
  \end{split}
\label{eq:product2}
\end{equation}
We notice that the sum $\sum_{i=0}^{L-1} w_j\smallsetminus v_i \equiv
\sum_{i=0}^{L-1} \mapping{A_j\smallsetminus B_i}{M_j\smallsetminus N_i}$ is not disjoint because every
term has in common the mapping:
\[
w_j\smallsetminus (\sum_{i=0}^{L-1} v_i) \equiv
\mapping{A_j\smallsetminus \cup_{i=0}^{L-1} B_i}{M_j \smallsetminus
  \cup_{i=0}^{L-1} M_i}
\]
also $w_j\smallsetminus v_0$ and $w_j\smallsetminus v_1$ have in
common the one above and $\sum_{i=2}^{L-1} w_j\cap v_i$, which are
already included in the second term in Equation
\ref{eq:product2}. Thus the first term in Equation \ref{eq:product2}
becomes basically $\sum_{j=0}^{K-1} w_j\smallsetminus
(\sum_{i=0}^{L-1} v_i)$.
\begin{equation}
  \begin{split}
    {\cal P}= &\sum_{j=0}^{K-1} w_j\smallsetminus (\sum_{i=0}^{L-1} v_i)+ \\
    & \sum_{j=0}^{K-1}\sum_{i=0}^{L-1}w_j\cap v_i +\\
    & \sum_{i=0}^{L-1} v_i\smallsetminus (\sum_{j=0}^{K-1} w_i)\\
  \end{split}
\label{eq:product3}
\end{equation}
Equation \ref{eq:product3} represents a disjoint mapping.

Let us return to Equation \ref{eq:partition1}.(part two) and
\ref{eq:p2refined} especially how to combine the terms that have
intersection: we can imagine that the index $j$ infers an order for
the components: the destination $L_0$, $L_1$, \dots , and
$L_{N-1}$. At the end of the first day $t_0$ we can summarize our
knowledge as
\begin{equation}
  \begin{split}
    D_{t_0} = & \sum_{i=0}^{N-1}\mapping{\dot{S}_{t_0}^{L_i}}{\dot{M}^{L_i}_{t_0}} +\\
             & \sum_{j=0}^{N-1} \prod_{i=0}^{N-1}\mapping{S^{L_i\rightarrow L_j}_{t_0}}{M^{L_i\rightarrow  L_j}_{t_0}}
  \end{split}
  \label{eq:partition-total}
\end{equation}

For the refinement in Equation \ref{eq:partition-total}, we have a
disjoint set of mappings. Now we must combine them. The first term in
Equation \ref{eq:partition1}.(part one) presents disjoint mappings and
thus can be just added in Equation \ref{eq:partition-total}. The
second term is a little trickier. We see it as the intersection of
mappings that have common users and thus narrowing the mapping
size. The second should reduce to a perfect matching and when it does
we can remove the users and put them {\em aside}.

Now let us consider the second day $t_1$. Let us compute $D_{t_1}$
independently from the previous step.  Then we join the two steps by
checking users intersections and refining the mappings: for each
mapping in $D_{t_1}$ we can make a product/intersection of each
mapping in $D_{t_0}$ and thus:
\[ D_{t_1} = D_{t_1} *D_{t_0}  \]

See the symmetric property of the product. Before any product or
update, the terms are a list of disjoint mappings. The product is meant
to combine mappings that have common addresses so that to refine the
mappings into matches. 

We should keep an order during the concatenation, for example:
{\tiny
  \begin{equation} 
    \begin{split}
      D_{t_{k+1}}=& \\
      D_{t_{k+1}} *D_{t_k} = & \Big( \sum_{i=0}^{N-1}\mapping{\dot{S}_{t_k}^{L_i}}{\dot{M}^{L_i}_{t_k}} +\sum_{j=0}^{N-1} \prod_{i=0}^{N-1}\mapping{S^{L_i\rightarrow L_j}_{t_k}}{M^{L_i\rightarrow L_j}_{t_k}}\Big) *\\ 
      & \Big(\sum_{i=0}^{N-1}\mapping{\dot{S}_{t_{k+1}}^{L_i}}{\dot{M}^{L_i}_{t_{k+1}}} +\sum_{j=0}^{N-1} \prod_{i=0}^{N-1}\mapping{S^{L_i\rightarrow L_j}_{t_{k+1}}}{M^{L_i\rightarrow L_j}_{t_{k+1}}}\Big) \\ 
    \end{split}
    \label{eq:computation}
  \end{equation}
}
\section{A Study in Parallelism}
\label{sec:parallelism}
The daily mappings $\mapping{S_{t_i}^{L_i}}{M_{t_0}^{L_i}}$ requires
the data from two consecutive days: $t_i$ and $t_{i+1}$. The first
parallel computation is based on the split of the interval of time
into smaller and consecutive intervals: two week interval each,
say. We compute each two-week interval in parallel. This is an
embarrassing parallelism.

\singlefigure{1.0}{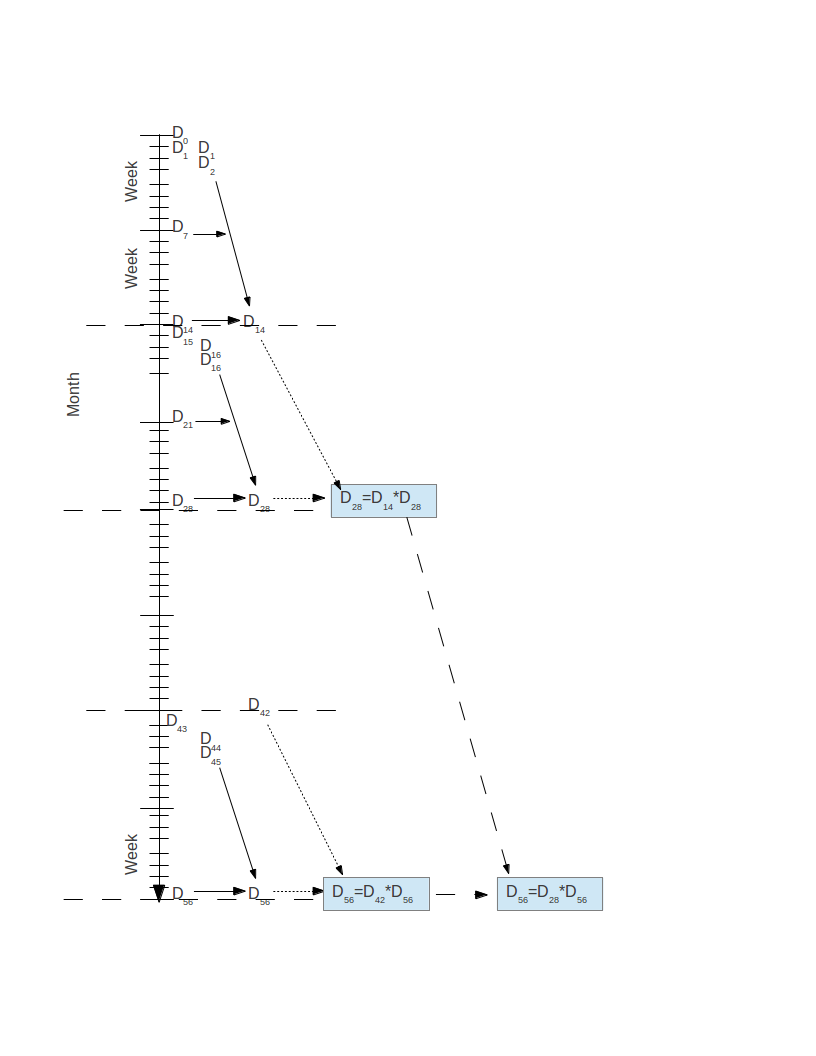}{Decomposition of the computation}{fig:figure1}

The total interval of time is composed of six months of data, we
actually split the computation into up to 15 independent
computations. Each $D_{t_j}$ is composed by a set of matches and
mappings. We take the list of $D_{t_j}$ and compute consecutive-pair
products as a binary tree.

Obviously, the last computation in the binary tree is a single product
and it seems that there is no parallelism to exploit. Take the example
in Figure \ref{fig:figure1}. The final product $D_{56} =
D_{28}*D_{56}$ will require at least as much as the sum of the
previous computations: $O(D_{14}*D_{28})+O(D_{42}*D_{56})$, which does
not seem parallel friendly.

In practice, as we go up in the tree, we loose explicit parallelism
but we can exploit the same amount of parallelism in the product.
Thus, we can keep the same level of parallelism throughout the
computation and thus efficient use of any architecture.  

The product becomes more complex as we go up.  In fact, the product
has to explore a Cartesian product of the operand mappings (graphs).
We explore if there are edges across the operands and this is why we
use the term adaptive for the graph we explore and build.

\section{The Implementation}
\label{sec:implementation}
{\tiny
\begin{verbatim}
# R implementation 
product <- function(D0,D1,P=2) {
  if (length(D0) ==0 && length(D1)==0)     {  R = list()   }
  else if ((is.null(D0) || length(D0) ==0) && length(D1)>0) {  R =  D1      }
  else if (length(D0) >0 && (is.null(D1) || length(D1)==0)) {  R =  D0      }
  else {
      L = group2(1:length(D0),length(D0)/P)
      ii <- function(K) {
          i=0; R = list(); D = list('S'=c(),'M'=c())
          for (k in K) {
              l = i
              Q = list('S'=c(),'M'=c())
              for (j in 1:length(D1)) { 
                  S = intersect(D0[[k]]$S,D1[[j]]$S)
                  M = intersect(D0[[k]]$M,D1[[j]]$M)
                  if (length(M)>0 && length(S)>0) {
                      i = i +1;  R[[i]] =list('S'=S,'M'=M)
                      Q$M = union(Q$M,M)
                      Q$S = union(Q$S,S)
                  }
              }
              if (i>l) { 
                  S = setdiff(D0[[k]]$S,Q$S)
                  M = setdiff(D0[[k]]$M,Q$M)
                  if (length(M)>0 && length(S)>0) {
                      i = i +1;  R[[i]] =list('S'=S,'M'=M)
                  }
                  D$M = union(D$M,Q$M)
                  D$S = union(D$S,Q$S)
              } else {
                  i = i +1;  R[[i]] = D0[[k]]
              }
          }
          list("R"=R, "D" = D, "disjoint"=(i==0)) 
      }

      RT = mclapply(L,ii,mc.preschedule=TRUE,mc.cores=P)

      R = list();   D = list('S'=c(),'M'=c());      disjoint = TRUE;   i=0
      for (rt in RT) {
          if (length(rt)>0) {
              for (r in rt$R) { 
                  i = i +1;  R[[i]] = r
              }
              disjoint = disjoint && rt$disjoint
              D$M = union(D$M,rt$D$M)
              D$S = union(D$S,rt$D$S)
          }
      }
      if (disjoint) {
          for (k in 1:length(D1)) {
              i = i +1;  R[[i]] = D1[[k]]
          }
      } else {
          for (k in 1:length(D1)) {
              S = setdiff(D1[[k]]$S,D$S)
              M = setdiff(D1[[k]]$M,D$M)
              if (length(M)>0 && length(S)>0) {
                  i = i +1;  R[[i]] = list('S'=S,'M'=M)
              }
          }
      }
      R
  }
}
\end{verbatim}
}

The product is the core of the whole computation. The software came
after the formal solution was found. The first implementation was
verbatim from Equation \ref{eq:product3}. This was a good starting
point. There are actually a few drawbacks: First, the intersections are
sparse and not balanced; that is, there may be intersection between
$S_i$ but not in between $M_i$ or viceversa. This means that the
computation $\sum_{j}\sum_{i}w_j v_i$ will spend quite some work
finding empty intersections and this information is not used for the
other terms.  Abusing a little the notation, we can rewrite the
computation in such a way that we avoid the graphs union computations
by computing the intersection first and reuse it:
{\footnotesize
  \begin{equation}
    \begin{split}
      T = & \emptyset \\
      {\cal P}= &
      \sum_{\ell=+\frac{K}{P}} 
      \Big( 
      \big(T{\cup=} W = \sum_{j=0}^{\frac{K}{P}} \sum_{i=0}^{L-1} w_{j+\ell} * v_i\big) + 
      \sum_{j=0}^{\frac{K}{P}} w_{j+\ell} \smallsetminus W 
      \Big) 
      +\\
      & \sum_{i=0}^{L-1} v_i\smallsetminus T\\
    \end{split}
    \label{eq:product4}
  \end{equation}
} Above, we present the implementation in R of the product. The
operand $D0$ is the mappings at time $t_i$ ($t_0$) and the operand $D1$
is the mappings at time $t_{i+1}$ ($t_1$).

The implementation has an important difference from Equation
\ref{eq:product3} and the intent in Equation \ref{eq:product4}, the
intersection has to be non empty for both $S$ and $M$ to be recorded
and used (in the following set difference computation). If we could
apply equation \ref{eq:product3}, the final product will be the
composition of {\em disjoint} terms. The implementation of Equation
\ref{eq:product4} does not assure that the final product has disjoint
terms and actually it may allow identical terms to appear, thanks to
the symmetric nature or the graph. Our implementation allows a minimum
and consistent computation: equal terms are removed and no-disjoint
terms involving perfect matches are simplified.

As we can see, the product explores all pairs to find intersections, a
square effect on the computational complexity.  Our implementation
choice is based on reducing the complexity even though by a constant.

\section{The Case Study}
\label{sec:case}
Due to the proprietary nature of the data, we cannot share the set
itself and a few of its details. However, we share the code verbatim
because of its simplicity (and will share the code upon request).

We observed about six airports for about six months.  We observed
about five hundred thousand unique MACs that appear more than once (if
there is only one appearance, there is very little signal and a
matching will be possible only if we match all other MACs, which is
unlikely).

We observed nine million unique users collected in a radius of one
mile from the airports requested center of interest and they appeared
more than two times during the entire period. On average, we have 2.5
million unique users and ten thousand unique MACs per day.

We build the mappings using two different granularities. See Figure
\ref{fig:figure1}. We use a granularity of two and four weeks to start
the computation. This is to cope with the {\em randomness} of the user
observation: We can only obtain a sample of the users UUID and their
appearance or their lack  affect the matches and their
products. Also the asymmetric nature of the product implementation
exemplified of Equation \ref{eq:product4} will make the resulting
graphs different. Otherwise, the graphs should be completely
deterministic, consistently built and eventually identical.

\begin{table*}
  \caption{Two/four-week mapping graph (above) and user/mac distribution (below)}
  \vspace{0.5cm}
  \label{tab:twoweeks}
  \centering
  \begin{tabular}{|l|r|r|r|r|}
    \hline
     weeks & matches & mappings & users covered & macs coverage \\
    \hline \hline 
    2 & 30667 & 130304 & 23448155 & 689828 \\
    4 & 33912 & 126650 & 24153536 & 686038 \\
    \hline
    \end{tabular}
  \vspace{0.5cm}
  \caption{Ratio user/mac distribution} \label{tab:fourweeks}
  \vspace{0.5cm}
  \begin{tabular}{|r|r|r|r|r|r|r|}
    \hline
    weeks & Min. &1st Qu. &Median &Mean &3rd Qu. &Max. \\
    \hline \hline 
    2&0.007 & 3.000 & 8.000 & 80.990 & 30.000 & 28690.000 \\ 
    4&0.012 & 3.000 & 8.000 & 87.940 & 32.000 & 27440.000 \\
    \hline
    \end{tabular}
\end{table*}

The computation time also may differ because of the different sparsity
of the mappings and their combinations. We present the results
separately and we conclude this section with a few considerations.

\subsection{The two and four week graphs}
The process follows the one presented in Figure \ref{fig:figure1}, we
start building day-by-day graph up to two/four weeks. Then, we build
the full graph.

Using the same number of resources, 16 cores for each computation, the
four week graph is a little faster (i.e., 2 hrs faster for a 4 days
computation from end-to-end), provides more matches, fewer mappings
but more redundancy. Notice that the two-week graph exploits more
parallelism and it will be faster if more resource could be used at
the beginning of the process. 

If we take a graph and we compute the ratio of the users number over
the MACs number in each mapping, we can summarize the graphs using
their distributions. We summarize their distribution in Table
\ref{tab:fourweeks}. In practice, the two-week graph has fewer
matches but the mappings tend to be more refined than the mappings in
the four-week graph, Table \ref{tab:twoweeks}.

We have not tried to combine the results (four and two weeks); it is
possible to take the graphs combine the matches and then compute the
product of the mappings. This is left as future investigation.

\subsection{Considerations}

The problem formulation and its notations were used to write a first
implementation: the first prototype was applied to a small graph after
a few weeks. The choice to write the solution in R was for the ease in
connecting to different databases where the data were available. The
simple semantic of the language fit the original formulation well.

As we increased the size of the graph, we decided to keep the original
solution, exploit the R parallelism, and to beef up the hardware (from
8-cores 32 GB machine to 32-cores 128GB). However, the square
complexity (i.e., O($N^2$) nodes of the graph) forced us to tune the
code and to relax the computation. We had to exploit parallelism in a
way that it is not natural to R.

We would suggest to chose a different environment or language to
exploit parallelism at loop level. 

\section{Conclusion}
\label{sec:conclusion}

To the best of our knowledge, the problem is novel because the
refinement of the mappings requires the intersection of two different
sets. There is no truth given a priori and thus there is no
learning. This is an example of graph mining. To the best of our
knowledge, our solution is novel as well. 

We provide a formal definition of the problem and its solution in
order to start a conversation. The formal statement actually have been
driving our problem presentation and solution. The desire of a well
defined formalism helped us freeing ideas by means of no ambiguity and
dangerous and misplaced intuitions.

As result, we have a solution that balances parallelism in an elegant
fashion as it unfolds during the computation. This parallelism is not
common and we wanted to share its application. This, in itself, could
be attractive to others.

\end{document}